\def\G1915{GRS~$1915$+$105$}
\def\X1550{XTE~J$1550$-$564$}
\def\J1655{GRO~J$1655$-$40$}
\def\eg{{\it e.g.} }
\def\etal{{\em et al. } }
\def\ie{{\em i.e. } }

\def\dd #1 {{\frac{\partial}{\partial #1}}}

\def\mt#1{#1}
\def\correct#1{#1}

\def\tom{\tilde\omega}
\def\cs2{c_{S}^2}

\def\ltsima{$\; \buildrel < \over \sim \;$}
\def\simlt{\lower.5ex\hbox{\ltsima}}
\def\gtsima{$\;\buildrel>\over\sim\;$}
\def\simgt{\lower.5ex\hbox{\gtsima}}

\documentclass{aa}
\usepackage{color}
\usepackage{graphicx}

\begin{document}

\title{Accretion-ejection instability and QPO in black-hole binaries.\\
II. Relativistic effects}

\author{P. Varni\`ere, J. Rodriguez  and M.Tagger}

\offprints{P.~Varni\`ere \\
(pvarni@discovery.saclay.cea.fr)
}

\institute{DSM/DAPNIA/Service d'Astrophysique (CNRS URA 2052), CEA 
Saclay, 91191 Gif-sur-Yvette, France}

\authorrunning{Varni\`ere, Rodriguez and Tagger}

\titlerunning{Accretion-Ejection Instability and QPO}

\date{Received 23 May 2000; accepted 25 February 2002}
\abstract{ The Accretion-Ejection Instability has been proposed to
explain the low frequency Quasi-Periodic Oscillation (QPO) observed in
low-mass X-Ray Binaries, in particular Black-Hole candidates.  Its
frequency, typically a fraction of the Keplerian frequency at the disk
inner radius, is exactly in the range indicated by observations.  The
variations of the frequency with the disk inner radius (extracted from
spectral fits of the X-ray emission) might thus be a useful test.  In
this paper we discuss how changes in the rotation curve, due to
relativistic effects when the disk approaches the central object, affect
\mt{the physics of the instability, and thus this frequency-inner
radius relation}.  We find that the relationship between the frequency
of the mode and the Keplerian frequency at the inner disk radius
($r_{int}$) \mt{departs from the one obtained in a Keplerian disk},
when $r_{int}$ approaches the last stable orbit.  This might agree with
the recently published results, showing a discrepancy between the
behavior of the QPO in the micro quasar \J1655, compared to other
sources such as \X1550 and \G1915.\\
In a companion paper (Rodriguez {\em et al.}, 2002, hereafter Paper I)
we have presented detailed observational results for \J1655 and \G1915. 
We show how the opposite correlations found in these sources between the
disk color radius (assumed to be close to its inner radius) and the QPO
frequency could indeed be explained by our theoretical result.
}
\maketitle
\section{Introduction}
\label{sec:Intro}
This is the second of two papers where we compare the properties of the
Accretion-Ejection Instability (AEI), recently found to occur in the
inner region of disks with a moderate magnetic field (Tagger and Pellat,
1999, hereafter TP99), with the low frequency Quasi-Periodic Oscillation
(QPO) observed in galactic Black-Hole binaries. \\
The AEI is a spiral instability, driven by magnetic stresses, of disks
threaded by a magnetic field of moderate (\ie near equipartition)
amplitude.  It belongs to the same family as galactic spirals, driven by
self-gravity (see Binney \& Tremaine, 1987 and references therein) or
the Papaloizou-Pringle instability (Papaloizou and Pringle, 1985),
driven only by pressure forces.  It is essentially the same spiral
instability found by Tagger \etal (1990); \mt{this instability was quite
weak and thus unlikely to be very efficient in an accretion disk, but} it
was shown in TP99 that a different physical process, the corotation
resonance (analyzed as a coupling with a Rossby wave in the disk) could
give it a more sizable growth rate.  Rossby waves are familiar in
planetary atmospheres, where their most spectacular manifestation is the
Great Red Spot of Jupiter.  They propagate in flows with a gradient of
vorticity, \mt{a gradient most often neglected in analytical studies of 
disks.}\\
In TP99 it was at the same time recognized that this resonance offers in
a magnetized disk a unique prospect, responsible for the name given to
the instability: the AEI grows (and causes accretion) by extracting
energy and angular momentum from the disk, and storing them in a Rossby
vortex at its corotation radius \mt{(the radius where the wave rotates
at the same velocity as the gas in the disk)}.  If the disk has a
low-density corona, this energy and momentum will be re-emitted as
Alfv\'en waves traveling {\em upward} to this corona, where they might
power a wind or a jet.  Thus, and although only a limited computation of
this effect was given in TP99 (a full computation will be presented in a
forthcoming paper), the AEI provides a unique way of connecting
accretion and ejection in the disk.  This contrasts with other known
disk instabilities, or with the hypothesis of a turbulent viscosity,
which all result in a {\em radial} transport of energy and momentum,
making a connection with MHD models of jets very difficult.  These
models show that the jet is very efficient at carrying away angular
momentum from the disk, and a mechanism connecting radial and vertical
transport of angular momentum is thus highly desirable (see however
Casse and Ferreira (2000) and references therein).  \mt{It is very
interesting to note\footnote{J. Ferreira, private communication} that
the conditions of instability are {\em always} fulfilled by these MHD
models of jets, since they have a plasma $\beta\sim 1$, and the radial
gradient, which must be positive, always has (as a consequence of the
self-similar ansatz) the value $+1/2$.}\\

In this paper we will first present a comparison between the observed
properties of the low-frequency QPO and the behavior expected from the
AEI, showing that this instability could provide a good explanation for
the QPO. \mt{In this respect we will turn to an effect neglected in
TP99.  In that paper it was found, from exact numerical solutions, that
the $m=1$ mode (\ie 1-armed spiral) was often the most unstable,
although it could not be predicted from an approximate WKB theory.  This
was confirmed in numerical simulations by Caunt and Tagger (2001).}
Here we first show that
relativistic effects, when the inner edge of the disk approaches the
Black Hole, change this special status of the $m=1$ mode, and introduce
a qualitative difference in its properties.  Numerical solution of the
linearized MHD equations, using a pseudo-Newtonian potential which
mimics the relativistic corrections to the rotation curve in the disk,
then shows that the relation between the mode frequency and the disk
inner radius is changed by these effects.  This is potentially important
since spectral fits of the disks of X-ray binaries give a measure of the
disk inner radius: thus the correlation between this radius and the QPO
frequency can be directly tested.\\
As we were completing this work which had started from a
purely theoretical argument, new observations (Sobczak {\em et al.},
2000, hereafter SMR)) showed that this frequency-radius correlation is
reversed in the microquasar \J1655 compared to other sources - in
particular another microquasar, \X1550.  We suggest that these
observations might correspond to our theoretical results.  In our
previous paper (Rodriguez {\em et al.}, 2001, hereafter paper I) we have
critically reassessed the analysis of \J1655 by SMR and confirmed its 
reverse frequency-radius correlation. We then turned to another 
microquasar, \G1915, because its high variability and the large number 
of available observations let us hope to explore the correlated 
variations of the disk radius and the QPO frequency over a much broader 
interval. The results for \G1915 showed the usual correlation, opposite 
to that of \J1655. In our final discussion we will show that these 
opposite behaviors might be explained by our theoretical result.\\

In a recent paper Psaltis and Norman (2001) have presented a model for
the low-frequency QPO. It is a model of random excitation of modes
(analogous in a way to the excitation of solar modes), which are
filtered at a transition radius assumed to exist in the disk.  The QPO
frequencies then lie at (or near) fundamental single-particle
frequencies at this radius.  In that case the low frequency QPO we
discuss here would be associated, as proposed by Stella and Vietri
(1999), with the nodal-precession frequency.  The computation we present
here is a more elaborate one (since it describes global hydrodynamical
or MHD perturbations, rather than the individual motion of isolated
particles or gas blobs), for one such type of mode (the spiral density
wave), which is based on the epicyclic motion.  It thus goes in the
direction outlined by Psaltis (2000), moving from basic frequencies of
motion to the hydrodynamical (in his case) or MHD (in our case)
properties of flows in the disk.  Our model is also different in that
the modes are {\em unstable}: they grow spontaneously from the thermal
noise and are expected (this has now been verified in numerical
simulations by Caunt and Tagger, 2001) to form narrow and high-amplitude
features, as galactic spirals do.  This contrasts with the model of
Psaltis and Norman, where a sharp transition radius must be assumed to
exist in the disk to explain the excitation of the mode, and we have no
{\em a priori} difficulty to explain the very high amplitude sometimes
observed in the QPO.
\section{The Accretion-Ejection Instability}
\label{sec:AEI}
\subsection{Basic properties}\label{subsec:basicAEI}
For the sake of completeness we present here the main elements of
physics underlying this instability mechanism.  More details and 
discussions can be found in TP99.
\newline
This instability appears as a spiral density wave in an accretion disk
threaded by a vertical (poloidal) magnetic field, of the order of
equipartition with the gas thermal pressure.  This part of the
instability is essentially similar to the spiral density waves of
galactic disks, but it is driven by magnetic stresses rather than
self-gravity (Tagger {\em et al.}, 1990).  The action of differential
rotation couples the spiral to a Rossby wave at its corotation radius
(the radius where the wave rotates at the same velocity as the gas). 
This means that the spiral wave generates, at its corotation radius, a
Rossby wave in which it stores the energy and angular momentum it
extracts from the disk (thus causing accretion).  It is this exchange of
energy and angular momentum which makes the whole perturbation (density
wave + Rossby wave) unstable, {\em i.e.} growing exponentially with
time.
\newline
\begin{figure*}
\centering   
\includegraphics[width=18cm]{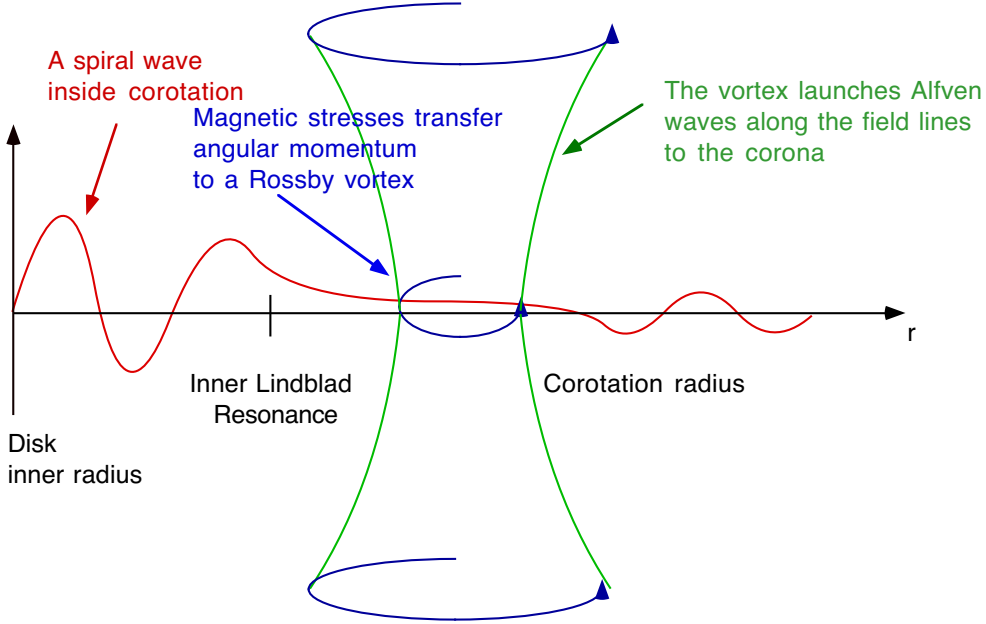} 
\caption{The structure of the instability is described here schematically as
a function of radius.  It is formed of a standing spiral density wave in
the inner part of the disk, coupled to a Rossby vortex it excites at its
corotation radius.  The Rossby vortex in turn generates Alfv\'en waves
propagating toward the corona of the disk.}
\label{fig:cavite}
\end{figure*}

A standing wave pattern (called in this context a normal mode) can form
in the following manner, also responsible for the standing spiral
pattern in galaxies: consider a spiral wave propagating outward from the
inner radius of the disk.  As it approaches corotation it is reflected
as another spiral wave, propagating inward; it is during this reflection
that some wave energy is exchanged (by the action of differential
rotation) with the Rossby wave, so that the reflected spiral has a
higher amplitude than the original one.  Now the reflected spiral, as it
reaches the inner disk radius, is reflected again as an outgoing spiral. 
This forms an equivalent to an electromagnetic cavity: if the resulting
outgoing spiral has the same phase as the initial one, the process will
repeat itself, the whole pattern being amplified at each reflection at
corotation.  As in an electromagnetic cavity, this integral phase
condition selects a discrete set of frequencies, corresponding to $n_{r}=0,\
1,\ 2\ldots$ nodes in the radial structure of the spiral wave pattern,
known as {\em normal modes} of the system.  In this process the Rossby
wave becomes a standing vortex, rotating at the angular velocity of the
pattern.  In practice only the $n_{r}=0$ mode, the most unstable one, will
concern us here.  Figure~\ref{fig:cavite} illustrates the trajectory of
the waves within the cavity, and the excitation of the Rossby wave.\\
This basic \mt{physical} description was obtained in the simple model of
a thin disk in vacuum, \mt{and numerical solutions were given in TP99}.  If
now one takes into account a low density corona above the disk, it was
shown in TP99 that the waves in the disk will generate an Alfv\'en wave
transferring to the corona (where it might power a wind or an outflow)
the energy and angular momentum extracted from the disk.  The AEI thus
appears as a very good candidate to connect accretion and ejection in
the inner region of a magnetized disk.  The computation of the Alfv\'en
wave emission was done, in TP99, in a WKB approximation valid only away
from the corotation region, where most of the emission should occur.  A
full computation (Varni\`ere and Tagger, 2002) will be given in a
forthcoming publication.
\subsection{The Inner Lindblad Resonance}\label{subsec:ILR}
The propagation properties of the waves determine the mode frequency,
and thus the location of its corotation radius.  It was shown in TP99
that the point where the wave is reflected, near its corotation radius,
is in fact the Inner Lindblad Resonance (ILR), where the Doppler-shifted
wave frequency
\[\tom(r)=\omega-m\Omega(r)\]
where $\omega$ is the wave frequency, \correct{$m$ is 
the azimuthal wavenumber (i.e. the number
of spiral arms)} and $\Omega$ is the rotation
(orbital) frequency, is equal to the opposite of the epicyclic frequency
$\kappa$, given by
\[\kappa^2=4\Omega^2+2\Omega\Omega'r\]
where the prime notes the radial derivative.  $\kappa$ is the restoring 
frequency which appears in the motion of individual particles (stars in 
the galactic context, fluid elements here), initially on a circular 
orbit, and to which a radial perturbation is given. When describing 
fluid motions, $\kappa^2$ appears as a restoring force, which 
supplements pressure and other forces (gravity in the galactic 
context, magnetic stresses here).

The wave propagates in the region where $\tom^2-\kappa^2>0$, \ie in the
radial interval between the inner disk radius $r_{int}$ and the ILR
radius.  This is where a cavity is formed, and where a standing pattern
results from the combination of the ingoing and outgoing waves, as shown
in figure \ref{fig:cavite}.  The manner in which waves are reflected at
the Inner Lindblad Resonance and at the inner disk edge has been
discussed in details in TP99. 

\mt{When studied in a WKB approximation for the radial structure of the
wave in a Keplerian disk, the wave is found to be evanescent 
(non-propagating) in the region between the ILR and the corotation radius.}
It was also shown in TP99 that \correct{in this WKB approximation} this
results in a mode frequency close to
\[\omega\simlt (m-1)\Omega_{int}\]
where $\Omega_{int}$ is the rotation frequency at the inner radius
$r_{int}$, and $m$ is the azimuthal wavenumber (\ie the number of spiral
arms).  \correct{From this WKB approximation} the $m=1$ mode seems to be
excluded, since it does not have an ILR because $\kappa=\Omega$ in
Keplerian rotation around a central mass (so that
$\omega-\Omega=-\kappa$ would imply $\omega=0$). However the WKB
analysis is only marginally valid, and should be used only as a guide. 
\correct{In practice it was shown in TP99 by exact numerical solution of
the problem that the $m=1$ mode does exist, and is often the most
unstable.} This was to be expected, since the physics of Rossby waves
usually selects large wavelengths ({\em e.g.} the Great Red Spot of
Jupiter), and since the efficiency of the corotation resonance (the
coupling between spiral and Rossby waves) is found to scale roughly as
$m^{-1}$.  Thus $m>1$ modes have their corotation radius very close to
$r_{int}$, whereas the $m=1$ typically has a frequency of the order of
$.1\ -\ .3\ \Omega_{int}$, depending on various disk parameters; this
corresponds to a corotation radius at a few times $r_{int}$.  As
discussed in the next section, this frequency range is the observed one
for the low frequency~QPO.\\

Relativistic effects may be expected to make a strong qualitative
difference here: indeed the last stable orbit at $r_{LSO}$ is defined as
the orbit where, because of these effects, the epicyclic frequency
$\kappa$ vanishes, whereas $\kappa=\Omega$ in a Newtonian, Keplerian
disk.  One must remember that $\kappa$ corresponds to the restoring
force experienced by a particle moving away from a circular orbit.  For
$r<r_{LSO}$, $\kappa$ is imaginary so that the radial motion of orbiting
particles is unstable and they will rapidly spiral toward the central
object.  The vanishing of $\kappa$ at $r_{LSO}$ \mt{(rather than
$\kappa=\Omega$ in a Newtonian, Keplerian disk)} thus introduces the
possibility for an $m=1$ mode to have an ILR, if the disk comes close
enough to $r_{LSO}$; \mt{we expect (and will indeed find below) this to
affect the properties of the $m=1$ mode.}\\

In the low state of black-hole binaries, $r_{int}$ is usually found to
be a few times $r_{LSO}$, although we do not know what can exist between
$r_{int}$ and the black hole: an ADAF might be a possibility, but one
might also think of the force-free magnetic structure necessary to
contain the vertical magnetic flux threading the black hole, according
to the Blandford-Znajek (1977) mechanism.  However certain observations,
which have been discussed in detail in paper I, show that sometimes,
during its variation, $r_{int}$ has a lower bound which probably marks
the position of $r_{LSO}$.  This has prompted us to study the effect of
the existence of an ILR on the frequency and growth rate of the $m=1$
mode.  We will then compare the theoretical result with the 
observational one of paper I.

\section{AEI and QPO properties}
\label{sec:compQPO}

Before taking into account the pseudo-Newtonian \mt{effects} we will
concentrate on how the AEI can be associated with the QPO. In this view
we will first make a brief summary of the characteristics of QPOs.

Quasi-Periodic Oscillations have been widely observed in many X-ray
binaries, whose compact object is either a neutron star or a black-hole
candidate.  They are commonly considered to originate in the disk,
either at its inner boundary (in particular the $kHz$ QPO in
neutron-star binaries) or beyond it.  Among black-hole binaries, the
low-frequency QPO (of the order of one to a few $Hz$) has drawn
particular attention, because it seems to convey important information
on the physics of the inner region of the accretion disk and of the
corona.

In particular in the micro-quasar \G1915, Swank {\em et al.} (1997) and
Markwardt {\em et al.} (1999) (hereafter SM97 for both of these
references), \mt{who dubbed it ``ubiquitous'' since it seems to be
always present in the low-hard state of the source,} have shown that its
frequency varies with the evolution of the disk, during its low and hard
state.  Indeed a correlation can be found, in two different manners:

\begin{itemize}
	\item The first correlation seems to relate the QPO frequency
	with the color radius of the disk.  This radius is obtained from
	the model (multicolor black-body+power-law tail) commonly used
	to fit the X-ray spectrum of the source.  \correct{It is
	considered as a measure of the inner radius of the disk, even
	though the exact relation between them is not well known.}
	\correct{We will discuss in this paper only in terms of the disk
	inner radius,} \mt{but will return below to the inner radius -
	color radius relation.}
	\\
    In the case of \G1915, during a typical 30 minutes cycle between a
    high/soft and a low/hard state, SM97 show that the QPO appears only
    during the low state, and that its frequency varies with the disk
    parameters.  If the QPO frequency is converted into an equivalent
    Keplerian radius in the disk, assuming a reasonable mass for the
    black hole, this radius would be of the order of a few times the
    observed color radius.  Furthermore, the QPO frequency seems very
    well correlated with the color radius.  Indeed, in a detailed
    discussion of the QPO in various states of the source, Muno {\em et
    al.} do show such a correlation.  In particular, during the 30 min. 
    cycles of \G1915, the QPO frequency can be seen to decrease with
    increasing radius.  Both studies conclude that the QPO, although it
    affects more strongly the power-law tail (coronal emission), seems
    to have its origin in the disk.\\

    \item A second correlation, found by Psaltis {\em et al.} (1999a)
    (hereafter P99), is more fragile at the present time since it relies
    on very few data for black-hole binaries.  Recent results by Nowak
    (2000), finding three broad peaks in the spectra of black-hole
    binaries, might indicate that there is more to be discovered in that
    direction.  The result of Psaltis {\em et al.} is very interesting
    because it fits, in a very different manner, with the previous one:
    they show that, in a large variety of X-ray binaries, there seems to
    be a correlation between a high and a low frequency QPO: the
    low-frequency QPO involved in the correlation \mt{(which is the QPO
    discussed by SM97 and by SMR)} is observed in many sources, whose
    compact object is either a neutron star or a black hole.  Its
    frequency ranges from about 1 to a few tens of Hz.  The
    high-frequency QPO in the correlation differs between neutron stars
    and black hole binaries: in neutron stars it is the lower of the
    pair of so-called ``kHz QPO'', believed to originate at the inner
    disk radius; the higher-frequency QPO of the pair would correspond
    to the Keplerian rotation frequency at this radius, while the
    lower-frequency one might be a beat wave between this and the
    rotation frequency of the neutron star (Miller {\em et al.}, 1998),
    or the periastron precession frequency (Stella and Vietri, 1999). 
    We will not discuss these models here (see {\em e.g.} Psaltis {\em
    et al.}, 1999b), but only retain that the upper frequency, in the
    correlation of Psaltis {\em et al.}, is close to the Keplerian
    rotation frequency near the inner edge of the disk.  In black-hole
    binaries, the high-frequency QPO considered in the correlation is
    seen as a broad feature at a few tens of Hz.  \correct{This
    correlation can thus be seen as relating the frequency of the
    low-frequency QPO and the inner radius of the disk.}
\end{itemize}
\mt{The correlation found by Psaltis {\em et al.} is such that the ratio
between the two frequencies is about 11, corresponding to a ratio~$\sim$
5 between the corresponding radii if one considers them as Keplerian
frequencies. This is comparable to the ratio found by SM97.}

	\correct{ The exact relation between the color radius given by
	the spectral fit and the {\em true} inner radius of the disk is
	not well known and an open debate among observers, \mt{since it
	involves a number of corrections which depend on assumptions
	about the disk and coronal structures}.  Different works explore
	this relation, such as Shimura \& Takahara (1995) in the case of
	a Schwarzschild black-hole, or Merloni, Fabian \& Ross (1999),
	but a general agreement on the correction to be made is still
	lacking.  \mt{In this work we will simply identify $r_{col}$ and
	$r_{int}$, and will discuss in our conclusions how this affects
	our results.} }
\newline

We consider the AEI as a very good candidate to explain the QPO because
of a number of characteristics:
\begin{itemize}
     \item Its frequency, for the one-armed spiral ($m=1$, where $m$ is
     the azimuthal wavenumber) is a fraction of the Keplerian frequency
     at the inner radius, fully compatible with the observation.  This
     would explain both the absolute value of the QPO frequency and its
     correlation with the inner disk radius.  We focus on the $m=1$ mode
     because the theory shows it (or the $m=2$, depending on disk
     parameters) to be most unstable, and because recent numerical
     simulations (Caunt and Tagger, 2001) show that the disk evolution
     most often leads it to dominate.  The $m=2$ would have a higher
     frequency than the low-frequency QPO we discuss here.\\

    \item As in galaxies, its physics should make it form long-lived,
    standing patterns, so that it should appear as what is called in the
    galactic context a quasi-stationary spiral structure - as confirmed
    by the numerical simulations.  This would explain the long life of
    the QPO (or rather its long correlation time, measured by the width
    of the QPO), as observed in particular during the hard and steady
    state of the source.\\

    \item These properties would probably be shared by any spiral mode,
    driven by other instability mechanisms ({\em e.g.} the
    Papaloizou-Pringle instability, although in that case the growth rate
    is weaker and concerns only high-$m$ modes).  However the AEI has
    the additional property of emitting {\em vertically}, as the
    Poynting flux of an Alfv\'en wave, the energy and angular momentum
    extracted from the disk (whence the name we have given to this
    instability); this makes it a realistic explanation, both for the
    compact jet observed in these sources ({\em e.g.} Dhawan {\em et
    al.}, 2000) and for the strong modulation of the coronal emission
    with the QPO. 
    
\end{itemize}

Based on these considerations, we have discussed (Tagger, 1999) how, if 
the QPO is indeed identified with the AEI, this could lead to a 
possible scenario for the 30 min. cycles of \G1915. In this scenario 
accretion in the inner region of the disk, and the observed cycles, 
would be controlled by the accumulation of poloidal magnetic flux in the 
disk.

These points may be considered only as favorable indications - though
better than for other models of QPO, where most often only basic
frequencies of motion are identified, without a mechanism of wave
excitation.  The basic physics and other theoretical expectations ({\em
e.g.} the formation of a long-lived, quasi-stationary spiral pattern)
have been confirmed by numerical simulations (Caunt and Tagger, 2001). 
This has led us to look for more distinctive observational signatures
of the AEI.

\section{Relativistic effects}
\label{sec:Pseudonewt}

A fully relativistic description of the instability is a
formidable challenge we will not attempt here.  Instead we use
existing models of Pseudo-Newtonian potentials, devised to mimic the
behavior of the rotation curve (and in particular the vanishing of
$\kappa$ at $r_{LSO}$) in the relativistic part of the disk.  This is
enough to check the effect, on the $m=1$ mode, of the apparition of an
Inner Lindblad Resonance in the disk.

\subsection{The Pseudo-Newtonian Potential}
\label{sec:potential}

We use the Pseudo-Newtonian potential given by 
Nowak \& Wagoner (1992):
\begin{eqnarray}
\Phi=-\frac{GM}{r}\left( 1-3\frac{GM}{rc^2}
+12\left(\frac{GM}{rc^2}\right)^2\right) \label{equ:pseudoN}
\end{eqnarray}
This is a second order approximation of the Schwarzschild metrics, so
that it neglects the effect of the spin of the compact object.  Although
the angular momentum arising from the spin of the black hole might
change the radius of the last stable orbit by a factor up to six for
extreme spin, the qualitative effect of the presence of an ILR for the
$m=1$ mode would remain.
\newline

\begin{figure}[htbp]
\centering   
\includegraphics[width=\columnwidth]{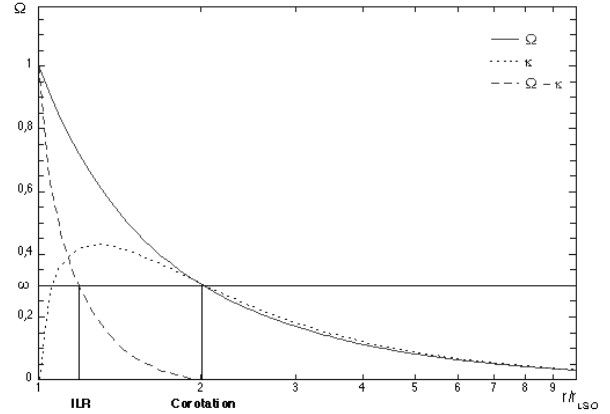}
\caption{Rotation curves
in a pseudo-Newtonian potential, showing the frequencies $\Omega$
(solid), $\kappa$ (dots) and $\Omega-\kappa$ (dashed), normalized to the
orbital frequency at the Last Stable Orbit ($r=r_{LSO}$).  The line at
$\omega=.3$ shows for example that a mode of frequency
$\omega=.3\Omega(r_{LSO})$ has a corotation and ILR respectively near
$r\simeq 2 r_{LSO}$ and $r\simeq 1.2 r_{LSO}$, if these radii are within
the disk ({\em i.e.} if the inner radius $r_{int}$ is small enough).  }
\label{fig:rotcurve}
\end{figure}

From this potential one can compute the orbital and epicyclic
frequencies:
\begin{eqnarray*}
    r\Omega^2 &=  & \frac{\partial \Phi}{\partial r } \\
    \kappa^2 & = & r\frac{\partial } {\partial r} \frac 1 r \frac{\partial \Phi} {\partial r} 
\end{eqnarray*}
so that the last stable orbit is at 
\[r_{LSO}=6\ \frac{GM}{c^2}\]
and $\Omega$ and $\kappa$ can be rewritten as:
\begin{eqnarray}
\Omega &=& \sqrt{GM} r^{-\frac{3}{2}} \left( 1- \frac{r_{LSO}}{r} 
	+ \left(\frac{r_{LSO}}{r}\right)^2 \right)^{\frac{1}{2}}\\
\kappa &=& \sqrt{GM}r^{-\frac{3}{2}}\left( 1- \left(\frac{r_{LSO}}{r}\right)^2 
	\right)^{\frac{1}{2}} 
\end{eqnarray}
Figure \ref{fig:rotcurve} shows the resulting rotation curve, and the
function $\Omega-\kappa$.  
\mt{It shows how relativistic effects (or a  pseudo-Newtonian 
potential) allow $\kappa$ to decrease near $r_{LSO}$, so that 
the $m=1$ mode can have an ILR when $r_{int}$ is close enough to $r_{LSO}$.}

For the numerical solution, we have also used as a cross-check the
Pseudo-Newtonian approximation of Paczynski \& Witta (1980):
\begin{eqnarray}
\Phi= -\frac{GM}{r-r_g}
\end{eqnarray}
where $r_g=2{GM}/{c^2}$ is the Schwarzschild radius.  This Pad\'e
approximation (rather than the series expansion above) gives essentially
the same results, confirming our expectation that the results depend
mostly on the qualitative difference introduced by the existence of an
ILR, rather than on the detailed rotation curve; we will present here
only results using equation \ref{equ:pseudoN}.
	
\section{Numerical Results}
\label{sec:num}

\subsection{Numerical Method}

We use the same method of solution as in TP99, and the same setup: the
disk is supposed to be infinitely thin and embedded in a vacuum.  It is
threaded by a vertical magnetic field $\vec B=B_{0}(r) \vec e_{z}$.  For
the sake of completeness, we repeat here the system of linearized MHD
equations solved:
\begin{eqnarray}
-i\tilde{\omega} U -2\Omega V &=& -c_s^2\frac{\partial h}{\partial s} 
-2\frac{B_0}{\Sigma} \frac{\partial \Phi_M}{\partial s} \label{equ:eulU}\\
-i\tilde{\omega} V +W U &=& -im c_s^2 h -2im\frac{B_0}{\Sigma} \Phi_M
 \label{equ:eulV}\\
-i\tilde{\omega}r^2 \sigma &=& -\frac{\partial}{\partial s} (\Sigma U)  
-im \Sigma V \label{equ:eulS}\\
-i\tilde{\omega}r^2 B_z^D &=&  -\frac{\partial}{\partial s} (B_0 U)
-im B_0 V \label{equ:eulB}
\end{eqnarray}
where $s=\ln r$, $U=rv_{r}$, $V=rv_{\vartheta}$, $\Sigma$ and $\sigma$
are the equilibrium and perturbed surface densities, $h=\sigma/\Sigma$,
$W=\kappa^2/2\Omega$ is the vorticity in the equilibrium flow, $B_{0}$
and $B_{z}^D$ are the equilibrium and perturbed vertical magnetic fields
at the surface of the disk, and $\Phi_{M}$ is a magnetic potential
obtained from:
\begin{equation}
    \nabla^2\Phi_{M}=-2B_{z}^1 \delta(z)
    \label{equ:laplPhi}
\end{equation}
which is solved using the Poisson kernel commonly used in
self-gravitating disks. In other terms, the magnetic field is described 
above the disk (in vacuum, with no currents) by a magnetic potential, 
whose source is the currents in the disk. For simplicity, since 
pressure plays very little role in the instability, we have assumed an 
isothermal equation of state.

The presence of $\Phi_{M}$ makes this an integro-differential system,
corresponding to the long-range action of magnetic stresses (in contrast
with the {\em local} action of pressure stresses between neighboring
fluid elements).  This system is solved by projecting it on a radial
grid, evenly spaced in the variable $s=\ln r$.  The solution then
reduces to finding the eigenvalues and eigenvectors of a matrix: each
eigenvalue is a frequency $\omega$ and the corresponding eigenvector
gives the radial structure of the perturbed velocity, density and
magnetic field.  A discrete set of eigenvalues, corresponding to the
modes with $n=0, \ 1, \ 2\ \ldots$ nodes in their radial structure, is
then easily identified, as discussed in TP99 (the other eigenvalues come
from the discretization of the problem on a numerical grid).  The
boundary condition of an outgoing wave at large radius ({\em i.e.} no
influx of information from the outer ranges of the disk) is implemented,
as in TP99, by solving on an axis slightly tilted in the complex-$s$
plane.

The only change here is that we use in equations
(\ref{equ:eulU}-\ref{equ:eulB}) the rotation curve described above,
rather than the Newtonian one used in TP99. We introduce the
parameter 
\[\xi={r_{int}}/{r_{LSO}}\]
At large $\xi$ the whole disk is very close to the Newtonian rotation
curve, and we recover the results of TP99.  When $\xi$ becomes close to
1, {\em i.e.} when the disk inner radius approaches the last stable
orbit, relativistic effects on the rotation curve start to play and the
$m=1$ mode can have an ILR. In that case, between $r_{int}$ and the ILR, the
wave propagates (in a WKB sense).  Let us compare this with the behavior
of a simple oscillator, described by the equation:
\[\frac{d^2}{d x^2}\Psi+M\Psi=0\]
(this comparison is quite relevant since, where the WKB approximation
applies, the system of equations (5-9) reduces to a system of this
form).

In the propagation zone ({\em i.e.} for us between $r_{int}$ and the
ILR) $M$ is positive, so that $\Psi$ oscillates.  On the other hand in
the ``forbidden band'' (for us, between the ILR and corotation, or
between $r_{int}$ and corotation if no ILR is present) $M$ is negative
and the mode has an exponential behavior \footnote{Note however that, as
described by Tagger {\em et al.} (1990), in our problem the behavior of
$\Phi_{M}$ in the forbidden band is algebraic rather than exponential;
this is due to the long-range action of magnetic stresses, {\em i.e.} to
the $r^{-m-1}$ dependence of an $m-$polar field.  It allows waves in the
inner cavity to couple quite efficiently, across the forbidden band, to
the outgoing wave and to the Rossby vortex.  This results in a stronger
instability than the Papaloizou-Pringle one of unmagnetized disks,
because in that case pressure forces result in a classical exponential
decay in the forbidden band.}.

Thus the change in the behavior of the solutions between $r_{int}$ and
the ILR will change the integral phase condition, which determines the
mode frequency.  This is true even though, as mentioned in the
introduction, the WKB approximation is not really valid, resulting in
the very fact that the $m=1$ mode exists even in a Newtonian potential. 
Our first goal is to quantify these effects.
\subsection{Numerical Results}
\label{sec:numres}
The computation uses two more parameters: the magnitude of the magnetic
field, and its radial profile.  The field is measured by the parameter
$\beta={2\mu_0 p}/{B^2}$, \ie the ratio of the thermal and magnetic
pressures.  As discussed in TP99, our instability occurs when $\beta$ is
of the order of 1 \mt{or lower} (whereas the magneto-rotational
instability (Balbus and Hawley, 1991, Chandrasekhar, 1960, Velikhov,
1959) occurs only for $\beta>1$).  We will present here only results
with $\beta=.5$, \mt{and complement them with results at $\beta=1$ for
comparison.}

In the same manner, we use as in TP99 a flat radial density profile, and
a magnetic field profile varying only over a limited radial range,
around the corotation radius: it was shown in TP99 that the amplification
of the mode depends only on the local gradient, at the corotation
radius, of the quantity $\kappa^2 \Sigma/2\Omega B_{0}^2$, and these
somewhat artificial profiles allow us an easy solution, by limiting the
global variation of equilibrium quantities across the numerical grid. 
The real part of the frequency depends of course on the global profiles. 
Here we want to consider only the effect of the relativistic
modifications of the rotation curve.  Thus we will stick to these simple
radial profiles.  We will report in a future publication results
obtained with a different but more elaborate method of solution; this
allows us to consider more general profiles but does not change
qualitatively the results presented here.

Keeping all these parameters constant, we solve the system of equations
(\ref{equ:eulU}-\ref{equ:laplPhi}) varying $\xi$, the ratio
$r_{int}/r_{LS0}$, from 1 to 100.  The results are shown in figure
\ref{fig:xi-w}.  As expected at large $\xi$ we recover the results
obtained in the Newtonian case, since the rotation curve is unaffected
by relativistic effects.  In these results the mode frequency $\omega$
varies as $r_{int}^{-3/2}$: since the problem has no scale length
besides $r_{int}$, $\omega$ is just proportional to $\Omega_{int}$, the
orbital frequency at the inner radius.
\begin{figure}[htbp]
\centering   
\caption{
\mt{Frequency of the $m=1$ mode, normalized to the orbital frequency at
the Last Stable Orbit for a Schwarzschild black hole, as a function of
$\xi=r_{int}/r_{LSO}$.  At large $\xi$, {\em .i.e} when the inner radius
of the disk is far away from the last stable orbit, the whole disk obeys
Newtonian dynamics and $\omega$ varies as $r_{int}^{-3/2}$.  At lower
$\xi$, relativistic effects modify the gas rotation curve near the inner
disk edge and change the behaviour of $\omega$: from $\xi\approx 10$ the
gradient softens gradually and the the correlation is reversed at
$\xi\approx 1.3$.
}
}
\label{fig:xi-w}
\end{figure}

 For $\xi<10$, small departures from this scaling become visible.
%
A strong difference appears when $\xi\simlt 1.3$: the correlation 
between $\omega$ and $r_{int}$ changes sign, so that now $\omega$ 
decreases as $\xi$ approaches $1$.

The limited numerical resolution does not allow us to see a marked
difference in the behavior of the eigenfunction (the radial dependence
of the perturbed quantities) near $r_{int}$, when $\xi$ approaches 1. 
Nevertheless, the values found for $\omega$ show that an ILR appears in
the disk, for $\xi<1.4$: this means that the inverse frequency-radius
correlation is associated with the presence of the ILR, introduced by
the relativistic effects on the rotation curve.  \mt{We show in figure
\ref{fig:2beta} the results obtained with $\beta=.5$ and $\beta=1.$,
normalizing this time for more clarity the mode frequency to the
rotation frequency at the inner radius.  At large $\xi$ the whole disk
is Newtonian and $\omega/\Omega_{int}$ becomes constant.  At $\beta=1$,
because the frequency is higher, the ILR appears in the disk only for
$\xi\simeq 1.1$, and the turnover of the frequency-inner radius
correlation is also shifted to lower $\xi$. As a result the left part 
of the curve, with a positive correlation due to relativistic effects, 
shows a much more limited variation of $\omega/\Omega_{int}$.}
\begin{figure}[htbp]
\centering   
\caption{
\mt{Frequency of the $m=1$ mode, normalized to the orbital frequency at
the inner radius, as a function of $\xi$ for $\beta=.5$ (solid) and
$\beta=1.$ (dashed).  In the latter case the frequency is higher, so
that an ILR appears in the disk only for $\xi\simlt 1.1$. As a result the
part of the curve with a positive slope, \correct{when normalized to the
last stable orbit obital frequency, is very small compared to the 
$\beta=0.5$ case. This motivates our presentation of the two cases normalized
to the inner radius rotation frequency which allows the comparison and 
shows a similar behavior.}}}
\label{fig:2beta}
\end{figure}

\mt{Our main result, the change in the correlation between $\omega$ and
$\xi$, is physically based on a qualitative difference in the rotation
curve, which allows the mode to have an ILR in the disk when $\xi$ is
close to 1.  We thus expect that this result should be quite robust and
persist with more detailed relativistic models, \eg taking into account
the Kerr metrics of a spinning black hole, or more realistic profiles in
the disk}.

\section{The observed correlation}
\label{observ}
As mentioned in the introduction, a QPO frequency-radius correlation,
for the low-frequency QPO of black-hole binaries, has already been
pointed out: in particular, for the micro-quasar \G1915, by SM97 during
a particular cycle of the source, and more generally by Muno {\em et
al.} who find that, among all observed properties, the QPO frequency is
well correlated with the disk inner radius.

On the other hand Sobczak {\em et al.} (2000) have recently shown that
two other black hole candidates, \X1550 and \J1655, have contrasting
behaviors: in \X1550 the QPO frequency decreases as the radius
increases, as seen in \G1915, while in \J1655 the correlation is
opposite.  This has led us to the hypothesis that these sources lie on
opposite sides of our theoretical curve, figure \ref{fig:xi-w}, \ie in
\J1655 the disk inner radius would be very close to the last stable
orbit, while in \X1550 (and apparently in \G1915) the disk would be
farther from the black hole.

In paper I we have critically reassessed the observational results of 
SMR on \J1655, taking into account the uncertainties on the relation 
between the observed color radius and the disk inner radius; we have 
then turned to \G1915 because its high variability let us expect to 
explore the theoretical curve over a broader radial range.

\subsection{Fit parameters}

\correct{The theoretical curve uses non-dimensional parameters, namely
$\xi=r_{int}/r_{LSO}$ and $\omega/\Omega(r_{LSO})$.  This will allow us
to compare on the same figure objects with different masses and spins. 
Before showing results from the comparison between the theory and
observation we will briefly present the formula and physical range for
the last stable orbit radius $r_{LSO}$ and the rotation frequency at
this radius $\Omega(r_{LSO})$.}

\correct{The general formula for the last stable orbit radius and the
rotation frequency at that orbit in the Nowak \& Wagoner
pseudo-Newtonian potential are}
\begin{eqnarray}
r_{LSO} &=& \frac{GM}{c^2} \left( 3 + A_2 \pm \sqrt{(3-A_1)(3+A_1 +2A_2)}
\right) \nonumber \\
\mbox{with:} \nonumber\\
A_1 &=& 1+ \sqrt[3]{1-a_\star^2} (\sqrt[3]{1+ a_\star} + 
\sqrt[3]{1- a_\star}) \nonumber \\
A_2 &=& \sqrt{3a_\star^2 + A_1^2}\nonumber\\
a_\star &=& \frac{J}{Mc} \nonumber\\
\mbox{and:} \nonumber\\
\Omega(r_{LSO})&=&\sqrt{\frac{GM}{r_{LSO}^3}}\nonumber
\end{eqnarray}
\correct{Where $J$ is the angular momentum of the black hole, and
$a_\star$ its specific angular momentum.}

\correct{These expressions can be normalized with the value for a one
solar mass black hole, namely \mt{the Last Stable Orbit} $r_\bullet=8.9\
km$ and \mt{rotation frequency} $\Omega_\bullet=1.4\ 10^4\ s^{-1}$.  We
note $\delta = 3 + A_2 \pm \sqrt{(3-A_1)(3+A_1 +2A_2)}$ and $m$ the
black hole mass in units of the solar mass,} \mt{giving:}
\begin{eqnarray}
r_{LSO}&=&\  \  \frac{\delta}{6} \  \  \  \  \  \  \  m\  \  r_\bullet \label{eq:rLSO}\\
\Omega(r_{LSO})&=& \left(\frac{6}{\delta}\right)^{3/2} \frac{1}{m}\  
	\Omega_\bullet\label{eq:OmegaLSO}
\end{eqnarray}
\correct{$\delta$ varies continously from $1$ for a prograde maximally
rotating Kerr black hole ($a_{\star}=+1$) to \mt{$6$ for $a_{\star}=0$,
\ie a Schwarzschild black hole} and $9$ for a retrograde maximally
rotating Kerr black hole ($a_{\star}=-1$).}

\correct{The black hole mass gives us the range of variation for the
last stable orbit radius.  Table \ref{tab:range} shows the allowed
ranges.  For \J1655, we have used the value 7 $M_{\odot}$, given by
Orosz \& Bailyn (1995).  \correct{For \G1915, we have used the value
$14\ M_{\odot}$ given by Greiner {\em et al.} (2001).} $a_{\star}=0.93$
is the value found by Zhang {\em et al.} (1997) for \J1655, given here
as an example of almost maximally rotating black hole.}

\begin{table}[htbp]
\begin{center}
\begin{tabular}{|l|c|c|}
\hline
spin            & GRO J$1655$ & GRS $1915$ \\
\hline
\hline
$a_{\star}=1$           & $10.4$      & $20.8$ \\
\hline
$a_{\star}=0.93$        & $21.9$      & $43.7$ \\
\hline
$a_{\star}=0$           & $62.3$      & $124.6$    \\
\hline
$a_{\star}=-1$          & $93.4 $     & $186.9$ \\
\hline
\end{tabular}
\caption{ \correct{Range of possible last stable orbit radius for \J1655
and \G1915 depending on the rotation parameter $a_{\star}$.} }
\label{tab:range}
\end{center}
\end{table}

\subsection{Results from fits}

We show in figure \ref{fig:bothcorrel} the observational results for
both sources (retaining only data points found valid by the analysis of
Paper I), fitted with the theoretical curve.

\begin{figure}[htbp]
\caption{\itshape{Plot of
the QPO frequency vs.  the radius for both sources, together with the
theoretical curve from TP99.  Both axes as normalized in such a way that
the plot is mass and spin independent.  \correct{All the data points are
taken from tables $1$ and $2$ of paper I.} }}
\label{fig:bothcorrel}
\end{figure}
The theoretical curve is the one shown in figure \ref{fig:rotcurve} for
a ratio of the gas thermal pressure to the magnetic pressure in the disk
$\beta=.5$, and the same radial profiles as in TP99.  \mt{Changing these
parameters changes the absolute value of the frequency but not the
overall behavior although, as discussed in section \ref{sec:numres}, a
value of $\beta\simgt 1$ would not allow us to reproduce the whole range of
frequency variation for \J1655.}

We have fit the data points by fixing, for each
source, a reference value for the radius and QPO frequency.  
\correct{The choice of value for these reference points are taken in the 
range allowed by the mass determination of the objects, see table \ref{tab:range}.}

\mt{In this range we fit the data curves by moving them solidly up and
down ({\em i.e.} fitting the fiducial value of the rotation frequency at
the Last Stable Orbit), and sideways ({\em i.e.} fitting the fiducial
value of $r_{LSO}$), \mt{which changes neither their shape nor their
variation with the ratio $r_{int}/r_{LSO}$.} We choose to do this
because the absolute relation between the observed $r_{col}$ and the
real $r_{int}$ cannot be constrained, given the uncertainties in the
model used for the spectral fits, and because our theoretical values
correspond to a fiducial choice of disk parameters.  However it is quite
remarkable that, identifying the absolute values of $r_{col}$ and
$r_{int}$, our best fit gives for \J1655 a spin parameter
$a_{\star}\simeq .946$ for a mass of 7 $M_{\odot}$ , whereas from a
different procedure Zhang \etal (1997) find $a{\star}=.93$.} \mt{For
\G1915 we obtain a spin of $0.9765$, compatible with the $0.998$ found
by Zhang {\em et al.} (1997).  These parameters are summarized in table
\ref{tab:recap}.\\} \mt{This agreement between independent estimates may
however be a pure coincidence, since both are model dependent: in
particular we have taken a ratio $r_{int}/r_{col}=1$; changing this
would change the determination of $r_{LSO}$, and thus ultimately of the
spin; Zhang \etal (1997) use a model of disk emission, as discussed in
the conclusions of paper I, and changing their hardening factor would
have a similar result.  Inversely Strohmayer (2001) finds a lower spin
value for \J1655, based on relativistic precession models (see \eg
Stella and Vietri, 1999).  The present work can thus be considered only
as one more contribution in this debate, attempting to constrain the
disk and fit parameters from different perspectives.}

\mt{ It is also noteworthy that, before the estimate of Greiner \etal
(2001) for the mass of \G1915 was available, we had left this parameter
free and the best fit gave us a mass of $\sim 15\ M_\odot$, very close
to the measured value, for an almost maximally ($0.99965$) rotating Kerr
black hole.  On the other hand, for \J1655 a value of $\beta\approx 1$,
giving a lower inner radius for the turnover of the frequency-radius
correlation, would not have allowed a satisfying fit because (as shown
in figure \ref{fig:2beta}) the theoretical range of variation of
$\omega$ is too small.\\}

\begin{table}[htbp]
\begin{center}
\begin{tabular}{|l||c|c|c|}
\hline
Object      & $r_{LSO}$ (km) &mass ($M_\odot$) & spin               \\
\hline
\hline
GRO J$1655$ &  $20.5$          &$7$         & $0.946$ ($0.93$)    \\
\hline
GRS $1915$  &   $34.5$         &$14$     & $0.9765$ ($0.998$) \\
\hline
\end{tabular}
\caption{ \correct{Results given by the fit of the data by the
theoretical curve.  The spin values in braket are from Zhang {\em et
al.} (1997)}  }
\label{tab:recap}
\end{center}
\end{table}
\mt{ Thus the agreement found from figure \ref{fig:bothcorrel} would
mean that in \J1655 the inner radius stays very close to the last stable
orbit, whereas in \G1915 (which is in a different spectral state) it is
already much larger when the QPO has appeared and the spectral fits
return reliable values of $r_{col}$.\\}
\mt{The two data points for \G1915 at $\xi\approx 1.2$ are puzzling: it
would be very tempting to place them on the left (growing) part of the
theoretical curve, as for \J1655.  A minor change in the fit parameters
would easily accomodate that.  We prefer to consider this as
inconclusive since the observational evidence is fragile: these points
are the first two in this cycle of the source, and are in fact obtained
{\em before} the dip (\ie the transition to the low-hard state) as
explained in paper I. They correspond to the phase, at the end of the
high state, when the inner radius starts moving away from the low value
(presumably very close to $r_{LSO}$) it has in this state.  We have
discussed in paper I how these points could in fact be subject to the
corrections of MFR. Furthermore, for the second (highest) of these
points we have also explained that the determination of the frequency is
not certain, and that it might in fact be a harmonic of the fundamental
frequency.}

\section{Discussion}\label{sec:discuss}
\mt{The Accretion-Ejection Instability appears as a good candidate to
explain the low-frequency QPO of black-hole binaries, for a number of
reasons: its frequency which lies in the observed range, its connection
with the corona and high-energy emission, the fact that it is an
instability, \ie does not need an {\em ad hoc} excitation mechanism, and
that the magnetic field configurations used for MHD models of jets obey
the instability criterion.}  
\correct{
In this paper we have shown that, if one accepts this identification of 
the AEI as the source of the QPO, we are able to understand
 an otherwise unexplained behavior.}
\mt{The theory predicts an observational signature, the
turnover of the relation between the QPO frequency and the inner radius
of the disk.  The relation between this radius and the color radius,
extracted from spectral fits, is still too uncertain to give more than
an indicative value to the fits we present.  It is however very
comforting that these fits, which also depend on assumptions on the
physical parameters in the disk, give results in surprisingly close
agreement with independent estimates.  More observations showing the
changing correlations (perhaps even in a single source at different
times) would be needed to confirm it.  These observations could extend
to neutron-star binaries, since Psaltis {\em et al.} (1999a) find a link
between the low-frequency QPOs in many different X-ray binaries.  Future
work will be dedicated to this, and to testing the suggestion we made in
Paper I that the anomalous color radius frequently found in spectral
fits of different sources (including the two studied here) might
indicate the presence of a spiral shock or hot point in the disk.  }

\bibliographystyle{plain}

\begin{thebibliography}{99}

\bibitem{Balbus}
Balbus, S.A., and Hawley, J.F., 1991, ApJ {\bf 376}, 214

\bibitem{Binney}
Binney, J. and Tremaine, S., 1987, {\it Galactic Dynamics}, 
Princeton University Press

\bibitem{BZ77}Blandford, R. D., and Znajek, R. L.,
1977, MNRAS {\bf 179}, 433B

\bibitem{Casse}
Casse, F. and Ferreira, J., 2000, A\& A {\bf 353}, 1115

\bibitem{Caunt}
Caunt, S. and Tagger, M., 2001, A\& A {\bf 367}, 1095

\bibitem{Chandrasekhar}
Chandrasekhar, S., 1960, Proc. Nat. Acad. Sci. {\bf 46}, 253

\bibitem{dhaw00} Dhawan, V., Mirabel, I.F. and Rodriguez, L.F., ApJ, {\bf 
543}, 373

\bibitem{G01} Greiner, J., Cuby, J.G., McCaughren, M.J., 2001, 
Nature, {\bf 414}, 522.

\bibitem{MST99}Markwardt, C. B., Swank, J. H., and 
Taam, R. E., 1999 ApJ  {\bf 513}, 37
		
\bibitem{MMil98} Miller, M.C., Lamb, F.K., and Psaltis, D., 1998, ApJ 
{\bf 508}, 791

\bibitem{MMR99} Muno, M.P., Morgan, E.H., and Remillard, R.A., 1999, 
ApJ {\bf 52}, 321

\bibitem{NW92} Nowak, M.A. and Wagoner, R.V., 1992, ApJ, {\bf 393},
607.

\bibitem{NW00} Nowak, M.A., 2000, {\it to appear in} MNRAS (Astro-ph/000523)

\bibitem{PW80} Paczynski, B. and Witta, P.J., 1980,A\&A, {\bf 88}, 23.

\bibitem{PP} Papaloizou, J. C. B. and Pringle, J. E., 1985, MNRAS {\bf 213}, 
799.

\bibitem{PBK99} Psaltis, D., Belloni, T.  and van Der Klis, M., 1999a, ApJ {\bf 520}, 262.

\bibitem{PWH99} Psaltis, D., Wijnands, R., Homan, J., Jonker, P.G., van
der Klis, M., Miller, M. C., Lamb, F. K., Kuulkers, E., van Paradijs,
J., and Lewin, W. H. G., 1999b, ApJ, {\bf 520}, 763.

\bibitem{PN00} Psaltis, D. and Norman, C. 2001, ApJ, submitted (astro-ph/0001391)

\bibitem{PS00} Psaltis,D., 2000, 
{\em submitted to} ApJ (astro-ph/0010316) 

\bibitem{ROD00}Rodriguez, J., Varni\`ere, P., Tagger. M., and 
Durouchoux, P., 2002, {\em to be published in} A\& A {\bf (paper I)}

\bibitem{S99}Sobczak, G.J., McClintock, J.E., Remillard, R.A., Cui, W.,
Levine, A.M., Morgan, E.H., Orosz, J.A., and Bailyn, C.D., 2000, {\em to
be published in} ApJ (astro-ph/9910519) ({\bf SMR}).

\bibitem{SV99} Stella, L., and Vietri, M., 1999, 
ApJ {\bf 524}, L63

\bibitem{Str} Strohmayer, T.E., 2001, ApJ {\bf 552}, L49 

\bibitem{S98} Swank, J., Chen, X., Markwardt, C., and Taam, R. , 1997,
proceedings of the conference "Accretion Processes in Astrophysics: Some
Like it Hot", held at U. Md., October 1997, {\em eds.} S. Holt and T.
Kallman

\bibitem{T99} Tagger, M., proceedings of the $5$th Compton Symposium,
Portshmouth (USA), 1999 (astro-ph/9910365)

\bibitem{T91} Tagger, M., Henriksen, R.N., Sygnet, J.F. and Pellat, R.,
1990, ApJ {\bf 353}, 654.

\bibitem{TP99} Tagger, M., and Pellat, R., 1999, A\&A, {\bf 349} 1003 
{\bf (TP99)}

\bibitem{V00} Varni\`ere, P., Tagger, M., 2001, {\it submitted to} A\& A

\bibitem{Velikhov}
Velikhov, E.P., 1959, Sov. Phys. JETP Lett. {\bf 9}, 995

\bibitem{Z97b} Zhang, S. N., Cui, W. and Chen, W., 1997, ApJ, {\bf 
482}, L155.

\end{thebibliography}

\end{document}